\documentclass[%
 aip,
 jmp,%
 amsmath,amssymb,
 preprint,%
]{revtex4-1}

\usepackage{graphicx}% Include figure files
\usepackage{dcolumn}% Align table columns on decimal point
\usepackage{bm}% bold math
\newcommand\ot{\otimes}

\newcommand\join{\vee}

\newcommand\ttt{{\text{\rm t}}}
\newcommand\spa{{\text{\rm span}}\,}

\def\a{\alpha }
\def\b{\beta }

\newtheorem{thm}{Theorem}
\newtheorem{prop}[thm]{Proposition}
\newtheorem{lemma}[thm]{Lemma}

\begin{document}

\preprint{AIP/123-QED}

\title%[Sample title]
{Existence of product vectors and their partial conjugates in a pair of spaces}% Force line breaks with \\
%\thanks{Footnote to title of article.}

\author{Young-Hoon Kiem}
% \altaffiliation%[Also at ]
  \affiliation%[Also at ]
 {Department of Mathematics and Institute of Mathematics,
 % and Institute of Mathematics,
 Seoul National University, Seoul 151-742, Korea}%Lines break automatically or can be forced with \\
\author{Seung-Hyeok Kye}%
% \email{Second.Author@institution.edu.}
\affiliation{Department of Mathematics and Institute of Mathematics,
 % and Institute of Mathematics,
 Seoul National University, Seoul 151-742, Korea}

\author{Jungseob Lee}
% \homepage{http://www.Second.institution.edu/~Charlie.Author.}
\affiliation{%
Department of Mathematics, Ajou University, Suwon 443-749, Korea%\\This line break forced% with \\
}%

\date{\today}% It is always \today, today,

\begin{abstract}
Let $D$ and $E$ be subspaces of the tensor product of the $m$ and
$n$ dimensional complex spaces, with co-dimensions $k$ and $\ell$,
respectively. In order to give upper bounds for ranks of entangled
edge states with positive partial transposes, we show that if
$k+\ell<m+n-2$ then there must exist a product vector in $D$ whose
partial conjugate lies in $E$.
%If $k+\ell
%>m+n-2$ then there may not exist such a product vector.
If
$k+\ell=m+n-2$ then such a product vector may
or may not exist depending on $k$ and
$\ell$.
\end{abstract}

%Use showkeys class option if keyword
\maketitle

\section{Introduction}

A simple tensor $x\otimes y$ in the tensor product space $\mathbb
C^n\otimes\mathbb C^m$ is said to be a {\sl product vector}. The {\sl partial conjugate} of a product
vector $x\otimes y$ is nothing but the product vector $\bar
x\otimes y$, where $\bar x$ is the vector whose entries are given
by the complex conjugates of the corresponding entries. The notion
of product vectors and their partial conjugates play key roles in
the theory of entanglement, which is one of the main research
topics of quantum physics in relation with possible
applications to quantum information and quantum computation
theory.

Let $M_n$ denote the $C^*$-algebra of all $n\times n$ matrices over the complex field. A positive semi-definite
matrix in $M_{mn}=M_n \otimes M_m$ is said to be {\sl separable} if it is a convex sum of rank one positive semi-definite
matrices onto product vectors in $\mathbb C^n\otimes \mathbb C^m$. A positive semi-definite matrix in $M_n\otimes M_m$
is said to be {\sl entangled} if it is not separable. The cone, denoted by $\mathbb V_1$,
of all separable ones coincides with the tensor product
$M_n^+\otimes M_m^+$ of positive cones, which is much smaller than $(M_n\otimes M_m)^+$, where $M_n^+$ denotes the cone of all positive
semi-definite matrices in $M_n$. So, entanglement consists of $(M_n\otimes M_m)^+ \setminus M_n^+\otimes M_m^+$.

If $A\in (M_n\otimes M_m)^+$ is a rank one
matrix onto a product vector $x\otimes y$ then the partial transpose
$A^\tau$ of $A$ is also positive semi-definite rank one
matrix onto the partial conjugate $\bar x\otimes y$, where the
{\sl partial transpose} of a block matrix in $M_n\otimes M_m$ is given by
$\left(\sum_{i,j}a_{ij}\otimes e_{ij}\right)^\tau
=\sum_{i,j}a_{ji}\otimes e_{ij}$.
Therefore, if $A\in M_n\otimes M_m$ is separable then its partial transpose $A^\tau$ is also positive semi-definite.
This gives us a simple necessary condition, called the PPT (positive partial transpose) criterion for separability, as was observed by
Choi \cite{choi82} and Peres \cite{peres}. Throughout this note, we denote by $\mathbb T$ the convex cone of all
positive semi-definite matrices in $M_n\otimes M_m$ whose partial transposes are also positive semi-definite:
$$
\mathbb T=\{A\in (M_n\otimes M_m)^+:  A^\tau\in (M_n\otimes M_m)^+\}.
$$
With this notation, the PPT criterion says that $\mathbb V_1\subset\mathbb T$.

When $n=2$, it was shown by Woronowicz \cite{woronowicz} that $\mathbb V_1=\mathbb T$ if and only if $m\le 3$. Especially,
he gave an explicit example of $A\in \mathbb T$ which is not separable in the case of
$M_2\otimes M_4$. This kind of block matrix is called
a {\sl PPTES} ({\sl positive partial transpose entangled state}) when it is normalized. The first example of PPTES in $M_3\otimes M_3$
was found by Choi \cite{choi82}.

Recall again that if $A\ge 0$ is of rank one onto a product vector then $A^\tau$ is onto its partial conjugate. Therefore, it is natural
to look at the range spaces of $A$ and $A^\tau$ to check the separability of $A$. The range criterion for separability
\cite{p-horo} tells us: If $A$ is separable then there exists a family $\{x_\iota\otimes y_\iota\}$ of product vectors
such that
$$
{\mathcal R}(A)=\spa \{x_\iota\otimes y_\iota\},\qquad
{\mathcal R}(A^\tau)=\spa \{\bar x_\iota\otimes y_\iota\}.
$$
This condition for a pair of subspaces also appears in characterization of faces of the cone $\mathbb T$ which induce
faces of $\mathbb V_1$ \cite{choi_kye}.
We refer to the book \cite{BZ}
for another criteria for separability as well as a systematic approach to the theory of entanglement.

A PPTES $A$ is said to be an {\sl edge PPTES}, or just simply an {\sl edge state}
if the face of $\mathbb T$ which has $A$ as an interior point contains no
separable one, which is equivalent to saying that there exists no product vector $x\otimes y\in {\mathcal R}(A)$
such that $\bar x\otimes y\in {\mathcal R}(A^\tau)$, as was introduced in Ref. \onlinecite{lkhc}. In other words, an edge state
is a PPTES which violates the range criterion in an extreme way. Since every PPTES is expressed as the convex sum of
a separable state and an edge state, it is crucial to
classify edge states to understand whole structure of
PPTES.

In order to classify edge states, we first have to know for which
pairs $(D,E)$ of subspaces of $\mathbb C^n\otimes \mathbb C^m$ there exists no product vector $x\otimes y$
such that
\begin{equation}\label{conj}
x\otimes y\in D,\qquad \bar x\otimes y\in E.
\end{equation}
An edge state $A$ is said to be of {\sl type} $(p,q)$ if
$\dim {\mathcal R}(A)=p$ and $\dim {\mathcal R}(A^\tau)=q$.
It is natural to classify edge states by their types as was tried in Ref. \onlinecite{sbl}.

The question of finding product vectors satisfying the condition
(\ref{conj}) had been considered in Ref. \onlinecite{2xn}
and \onlinecite{hlvc} to distinguish separable states
among PPT states. It had been shown \cite{2xn} that
if $m=2$ and
$$
\dim D^\perp+\dim E^\perp< n=(2+n)-2
$$
then there exist infinitely many product vectors
with (\ref{conj}), and the separability of
PPT states had been discussed in the case
$\dim D^\perp+\dim E^\perp=n$.
The general cases
$$
\dim D^\perp+\dim E^\perp=(m+n)-2
$$
had been also discussed \cite{hlvc}
without definite conclusion on the existence
itself of product vectors with (\ref{conj}).
Just referring to these two papers, the authors of
Ref. \onlinecite{sbl} claimed the following:

\medskip
{\bf Claim:}
In the case of $m=n=3$,
if there exists an edge state of type $(p,q)$,
then $p+q <14$.
\medskip

This paper is an outcome of trying to understand this claim and find
conditions on the dimensions of $D$ and $E$ for which the existence
of a pair $(x\ot y,\bar x\ot y)\in D\times E$ is guaranteed. Those cases
are naturally excluded in the classification of edge states by their types.
Main results of this paper are listed in the following:
\begin{enumerate}
\item[(i)]
If $\dim D + \dim E > 2mn-m - n + 2$, then there exists a
pair $(x\ot y,\bar x\ot y)\in D\times E$.
\item[(ii)]
If $\dim D + \dim E = 2mn-m - n + 2$ and
\begin{equation}\label{int-eq}
\sum_{r+s=m-1}(-1)^r \binom kr\binom \ell s \neq 0
\end{equation}
with $k=\dim D^\perp$ and $\ell=\dim E^\perp$,
then there exists a
pair $(x\ot y,\bar x\ot y)\in D\times E$.
\item[(iii)]
If $\dim D + \dim E < 2mn - m - n + 2$, then such a pair is not
guaranteed to exist.
\end{enumerate}

By the first result (i), we have an upper bounds for the ranks
of edge states and their partial transposes
in terms of their types: If there is an
$m\otimes n$ edge state of type $(p,q)$ then
\begin{equation}\label{iugiuhil}
p+q \le 2mn-m-n+2.
\end{equation}
This upper bound may be known to specialists,
even though it is not proved explicitly in the literature.
Our proof involves binomial coefficients as well as
techniques from algebraic geometry.

Our main contribution is on the case of
$p+q = 2mn-m-n+2$. In this case, the second result (ii) tells us that
if (\ref{int-eq}) holds then there exists no edge state of type $(p,q)$.
This means that the equality
may be deleted in (\ref{iugiuhil}) for some $(p,q)$,
and it gives us more precise upper bounds than (\ref{iugiuhil})
for the existence of edge states.

In the $3\otimes 3$ case, we have $2mn-m-n+2=14$,
and it turns out that the pair $(k,\ell)=(2,2)$ satisfies the condition (\ref{int-eq}). This means
that there is no edge state of type $(7,7)$, and
the Claim is confirmed for $(p,q)=(7,7)$.
The pair $(k,\ell)=(1,3)$ does not satisfy the condition (\ref{int-eq}). In fact, we construct a pair
$(D,E)$ of subspaces with $\dim D=1$ and $\dim E=3$
for which there exists no pair product vector $x\ot y\in D$ with $\bar x\ot y\in E$.
Unfortunately, we cannot prove or disprove the existence
of an edge state $A$ with ${\mathcal R}A=D$ and
${\mathcal R}A=E$. The existence of $3\otimes 3$
edge states of type $(6,8)$ seems to be still open.
Recently, it is also claimed in Ref. \onlinecite{lein} that
if $D=({\mathcal R}A)^\perp$ and $E=({\mathcal R}A^\tau)^\perp$
for a PPT state $A$ and $\dim D+\dim E=m+n-2$ then there
exist finitely many pairs $(x\ot y,\bar x\ot y)\in D\times E$.
Our example shows that this is not true for general pairs
$(D,E)$ with the same dimension condition.

In the $2\otimes 4$ case, $(k,\ell)=(1,3)$ also satisfies
the condition (\ref{int-eq}), which means that there is no
edge state of type $(5,7)$. This special case was already proved in
Ref. \onlinecite{skl}.
By the same reason as in the case of $3\otimes 3$, we could not
determine the existence of an edge state of type $(6,6)$.

In the next section, we state and prove the main theorem mentioned above.
In Section 3, we analyze some exceptional cases
for which $p+q = 2mn-m-n+2$ but (\ref{int-eq}) does not hold,
and find explicit examples of pairs $(D,E)$ of subspaces without
pair $(x\ot y,\bar x\ot y)\in D\times E$,
in the case of $m=2$ or $m=n=3$. We close this paper
reviewing known examples of edge states
with various types in low dimensions,
and comparing the results on the existence of
product vectors in a single space.

%%%%%%%%%%%%%%%%%%%%%%%%%%%%%%%%%%%%%%%%%%%%%%%%%%%%%%%%%%
\section{Results}

\def\CC{\mathbb{C} }
\def\PP{\mathbb{P} }
\def\ZZ{\mathbb{Z} }
\def\cO{\mathcal{O} }

We begin with the following.

\begin{thm}\label{thm-l1}
Let $(k,\ell,m,n)$ be a quadruplet of natural numbers with the relation $k,\ell\le m\times n$. If
\begin{equation}\label{eq-l2}
(-\alpha+\beta)^k(\alpha+\beta)^l\neq 0\qquad\text{modulo} \quad \alpha^m,\beta^n,
\end{equation}
in the polynomial ring $\mathbb{Z}[\alpha,\beta]$,
then for any pair $(D,E)$ of subspaces of $\mathbb C^n\otimes\mathbb C^m$ with $\dim D^\perp=k$ and $\dim E^\perp=\ell$
there exists a nonzero product vector $x\ot y\in D$ with $\bar x\ot y\in E$.
\end{thm}
Precisely speaking, \eqref{eq-l2} means that $(-\alpha+\beta)^k(\alpha+\beta)^l$
is not contained in the ideal generated by $\alpha^m$ and $\beta^n$.
This is an application of intersection theory in algebraic geometry for which Ref. \onlinecite{Fulton} is a standard reference.

\medskip
{\bf Proof:}
Let $\PP(\CC^m\otimes \CC^n)$ denote the projective space of lines in $\CC^m\otimes \CC^n$.
Obviously, the locus of product vectors is the image of the Segre map
\[ \PP^{m-1}\times \PP^{n-1}\hookrightarrow \PP(\CC^m\otimes \CC^n)\]
defined by $([x],[y])\mapsto [x\otimes y]$, where $[x]$
(respectively $[y]$) denotes the line spanned by a nonzero vector $x$ (respectively $y$).

The integral cohomology ring of $\PP^{m-1}\times \PP^{n-1}$ is perfectly understood (see any basic textbook on algebraic topology) as
\[ H^*(\PP^{m-1}\times \PP^{n-1})\cong \ZZ[\alpha,\beta]/(\alpha^m,\beta^n) .\]
Let us define a homeomorphism
$$
\phi:\PP^{m-1}\times \PP^{n-1}\to \PP^{m-1}\times \PP^{n-1},\quad \phi([x],[y])=([\bar{x}],[y]).
$$
The induced isomorphism in cohomology is given by
$$
\alpha\mapsto -\alpha,\qquad
\beta\mapsto \beta
$$
since the orientation of a line in $\PP^{m-1}$ is changed.
Since the hyperplane bundle $\cO(1)$ over $\PP(\CC^m\otimes \CC^n)$
restricts to $\cO(1,1)$ on the product $\PP^{m-1}\times \PP^{n-1}$,
a general subspace in $\PP(\CC^m\otimes \CC^n)$ of codimension $k$
intersects with $\PP^{m-1}\times \PP^{n-1}$ along a cycle whose
Poincar\'e dual is
$(\alpha+\beta)^k$
in $H^*(\PP^{m-1}\times \PP^{n-1})\cong \ZZ[\alpha,\beta]/(\alpha^m,\beta^n)$.

Let $D$ (respectively $E$) be a subspace of codimension $k$ (respectively $\ell$)
in $\CC^m\otimes \CC^n$.
By the definition of $\phi$, it is obvious that there exists a
product vector $x\otimes y\in D$ with $\bar{x}\otimes y\in E$
if and only if $$\phi(\PP D\cap (\PP^{m-1}\times \PP^{n-1}))\cap \PP E\ne \emptyset.$$
By the standard intersection theory, small
perturbations $\Gamma_1,\Gamma_2$ of $\PP D$ and $\PP E$ give
us a transversal intersection $\phi(\Gamma_1\cap (\PP^{m-1}\times \PP^{n-1}))\cap \Gamma_2$
whose Poincar\'e dual is precisely
\begin{equation}\label{poly}
(-\alpha+\beta)^k(\alpha+\beta)^\ell
\end{equation}
in $H^*(\PP^{m-1}\times \PP^{n-1})\cong \ZZ[\alpha,\beta]/(\alpha^m,\beta^n)$.
Hence if
$(-\alpha+\beta)^k(\alpha+\beta)^l\neq 0$, then
$$
\phi(\Gamma_1\cap (\PP^{m-1}\times \PP^{n-1}))\cap \Gamma_2\neq \emptyset
$$
which in turn implies
$$
\phi(\PP D\cap (\PP^{m-1}\times \PP^{n-1}))\cap \PP E\neq \emptyset,
$$
since a small perturbation of empty intersection is still empty. This completes the proof.
$\Box$
\medskip
%\end{proof}

We expand the polynomial (\ref{poly}) to write
$$
(-\a +\b)^k(\a+\b)^\ell
=\sum_{t=0}^{k+\ell} C^{k,\ell}_t \a^t \b^{k+\ell-t}
$$
with the coefficients
$$
C^{k,\ell}_t=\sum_{r+s=t}(-1)^r \binom kr\binom \ell s.
$$
%If $k=0$ or $\ell=0$ then the polynomial (\ref{poly}) is always nonzero.
If $k+\ell=m+n-2$ then we have
$$
(-\a+\b)^k(\a+\b)^\ell
=\cdots
+C^{k,\ell}_{m-2}\a^{m-2}\b^{n}
+C^{k,\ell}_{m-1}\a^{m-1}\b^{n-1}
+C^{k,\ell}_{m}\a^{m}\b^{n-2}+\cdots.
$$
We see that the polynomial (\ref{poly}) is zero modulo $\a^m$ and $\b^n$
if and only if $C^{k,\ell}_{m-1}=0$. To deal with the case $k+\ell <m+n-2$, we need the following:

\begin{lemma}\label{zero-zero}
Let $k,\ell$ be nonnegative integers. When we expand the polynomial
$$
P^{k,\ell}(x) = (1-x)^k (1+x)^\ell
$$
and sort by degrees,
two consecutive coefficients of $P^{k,\ell}(x)$ cannot be zeros.
\end{lemma}

{\bf Proof:}
First of all, we have
$P^{k,\ell}(x)=\sum_{t=0}^{k+\ell} C^{k,\ell}_t x^t$.
As for the coefficients, we have the following identities
\begin{equation}\label{njjkkl}
\begin{aligned}
C^{k,\ell}_t  &= C^{k-1,\ell}_{t} - C^{k-1,\ell}_{t-1} \\
C^{k,\ell}_t  &= C^{k,\ell-1}_{t} + C^{k,\ell-1}_{t-1} \\
t C^{k,\ell}_t &= -k C^{k-1,\ell}_{t-1} + l C^{k,\ell-1}_{t-1}.
\end{aligned}
\end{equation}
The first and second identities immediately follow from the identities
$$
\binom{k}{r} = \binom{k-1}{r} + \binom{k-1}{r-1},\qquad
\binom{\ell}{s} = \binom{\ell-1}{s} + \binom{\ell-1}{s-1},
$$
respectively.
To prove the third one, we differentiate $P^{k,l}(x)$:
$$
\begin{aligned}
\frac{d P^{k,\ell}}{dx}(x) &= -k(1-x)^{k-1} (1+x)^\ell + \ell (1-x)^k(1+x)^{\ell-1} \\
&= -k P^{k-1,\ell}(x) + \ell P^{k,\ell-1}(x).
\end{aligned}
$$
On the other hand, we also have
$$
\frac{d P^{k,\ell}}{dx}(x) = \sum_{t=1}^{k+\ell} tC^{k,\ell}_t x^{t-1},
$$
from which the third identity follows.

Assume that $C^{k,\ell }_t = C^{k,\ell }_{t+1}=0$. Then by (\ref{njjkkl}), we have
\begin{eqnarray}
\label{1st} C^{k-1,\ell }_{t} - C^{k-1,\ell }_{t-1} &=& 0 \\
\label{2nd} C^{k,\ell -1}_{t} + C^{k,\ell -1}_{t-1} &=& 0  \\
\label{3rd} -k C^{k-1,\ell }_{t-1} + \ell C^{k,\ell -1}_{t-1} &=& 0 \\
\label{4th} C^{k-1,\ell }_{t+1} - C^{k-1,\ell }_{t} &=& 0 \\
\label{5th} C^{k,\ell -1}_{t+1} + C^{k,\ell -1}_{t} &=& 0 \\
\label{6th} -k C^{k-1,\ell }_{t} + \ell C^{k,\ell -1}_{t} &=& 0.
\end{eqnarray}
From equations (\ref{1st}), (\ref{2nd}) and (\ref{3rd}), we get
$k C^{k-1,\ell }_t + \ell C^{k,\ell -1}_t=0$.
This together with the relation (\ref{6th}) implies that
$$
C^{k-1,\ell }_t = C^{k,\ell -1}_t =0.
$$
On putting these into  (\ref{1st}), (\ref{2nd}), (\ref{4th}) and (\ref{5th}), we see that
$$C^{k-1,\ell }_{t-1} = C^{k,\ell -1}_{t-1} =C^{k-1,\ell }_{t+1} = C^{k,\ell -1}_{t+1}=0.$$
By induction this leads to a contradiction.
$\Box$
\medskip
%\end{proof}

By Lemma \ref{zero-zero}, it is immediate that if $k+\ell <m+n-2$ then the polynomial (\ref{poly}) is never zero
modulo $\a^m$ and $\beta^n$.
We summarize as follows:

\begin{thm}\label{cor-l1}
Let $m$ and $n$ be natural numbers, and $(k,\ell)$ a pair of natural numbers with $k,\ell\le m\times n$.
Consider the following condition:
\begin{itemize}
\item[{\rm (C)}]
For any pair $(D,E)$ of subspaces of $\mathbb C^n\otimes \mathbb C^m$ with
$\dim D^\perp =k$, $\dim E^\perp=\ell$, there exists
a nonzero product vector $x\otimes y\in D$ with $\bar x\ot y\in E$.
\end{itemize}
Then we have the the following:
\begin{enumerate}
\item[{\rm (i)}]
If $k+\ell>m+n-2$ then the condition {\rm (C)} does not hold.
\item[{\rm (ii)}]
If $k+\ell<m+n-2$ then the condition {\rm (C)} holds.
\item[{\rm (iii)}]
In the case of $k+\ell=m+n-2$, if $C^{k,\ell}_{m-1}\neq 0$ then the condition {\rm (C)} holds.
\end{enumerate}
\end{thm}

{\bf Proof:}
The statements (ii) and (iii) are direct consequences of Theorem \ref{thm-l1}.
The statement (i) is obtained by a dimension count:
Let $Gr(mn,k)$ be the set of all subspaces of $\CC^n\otimes\CC^m$ of codimension $k$.
Then it is easy to see that $Gr(mn,k)$ is a manifold of dimension $k(mn-k)$
since the tangent space at $D\in Gr(mn,k)$ is $\mathrm{Hom}(D,D^\perp)$.
Using the notation of the proof of Theorem \ref{thm-l1}, the condition
(C) holds if and only if
$$
\phi(\PP D\cap (\PP^{m-1}\times \PP^{n-1}))\cap \PP E\ne \emptyset
$$
for all subspaces $D$ and  $E$ of co-dimensions $k$ and $\ell$ respectively.
By Bertini's theorem\cite{Hartshorne}, if we choose a general
$D$, $\phi(\PP D\cap (\PP^{m-1}\times \PP^{n-1}))$ is a connected manifold
of real dimension $2m+2n-2k-4$. For each point $p$ in $\phi(\PP D\cap (\PP^{m-1}\times \PP^{n-1}))$,
the set of $E\in Gr(mn,\ell)$ containing $p$ is diffeomorphic to
$Gr(mn-1,\ell)$ because it suffices to choose a codimension $\ell$ subspace
in the quotient of $\CC^n\otimes \CC^m$ by the line of $p$. Since the
real dimension of $Gr(mn-1,\ell)$ is $2\ell (mn-\ell-1)$, varying $p$ in
$\phi(\PP D\cap (\PP^{m-1}\times \PP^{n-1}))$, we obtain a manifold (actually a fiber bundle) of dimension at most
$$
(2m+2n-2k-4)+ 2\ell (mn-\ell-1)
=2\ell (mn-\ell ) +2(m+n-2-k-\ell)
$$
which is smaller than the real dimension $2\ell (mn-\ell)$ of $Gr(mn,\ell)$
when $k+\ell>m+n-2$. Therefore, for a general choice of $D$, the set of $E$
for which there exists a nonzero product vector $x\otimes y$ with \eqref{conj} is a
proper subset in $Gr(mn,\ell)$. This obviously is sufficient for the statement (i).
$\Box$
\medskip
%\end{proof}

%%%%%%%%%%%%%%%%%%%%%%%%%%%%
\section{exceptional cases and examples}

The only remaining \lq exceptional\rq\ cases are when
the relation
\begin{equation}\label{eq-l3}
k+\ell=m+n-2 \quad \text{and}\quad C_{m-1}^{k,\ell}=0
\end{equation}
holds.
We note that the first equation of \eqref{eq-l3} denotes just the green lines in the figures of Ref. \onlinecite{lein}
in the context of PPT states.
We consider several easy
cases when the relation (\ref{eq-l3}) hold in the following proposition. The proofs will be omitted.
\begin{prop}\label{relation}
We have the following:
\begin{enumerate}
\item[{\rm (i)}]
When $m=2$, the relation \eqref{eq-l3} holds if and only if
$n=2k$ and $\ell=k$.
\item[{\rm (ii)}]
When $m=3$,  the relation \eqref{eq-l3} holds if and only if
$$
n=r(r+2),\quad k=\binom{r+1}{2}\quad \text{and} \quad \ell=\binom{r+2}{2}
$$
for a positive integer $r$.
\item[{\rm (iii)}]
When $m=n$,  the relation \eqref{eq-l3} holds if and only if $k$ and $\ell$ are odd.
\item[{\rm (iv)}]
When $k=\ell$,  the relation \eqref{eq-l3} holds if and only if $m$ and $n$ are even.
\end{enumerate}
\end{prop}

Let $Gr(mn,k)$ (respectively $Gr(mn,\ell)$) denote the set of all subspaces of $\mathbb C^n\otimes\mathbb C^m$
of codimension $k$ (respectively $\ell$), as in the proof of Theorem \ref{cor-l1}. We denote by
$A(m,n,k,\ell)$
the set of all $(D,E)\in Gr(mn,k)\times Gr(mn,\ell)$ such that there exists a nonzero product vector $x\otimes y$
satisfying (\ref{conj}). Then $A(m,n,k,\ell)$ is a proper
subset of $Gr(mn,k)\times Gr(mn,\ell)$ if and only
if there exist subspaces $D$ and $E$ of co-dimensions
$k$ and $\ell$ respectively for which there exists no
nonzero product vector $x\otimes y$ satisfying (\ref{conj}).
By Theorem \ref{cor-l1}, $A(m,n,k,\ell)$ equals the
whole set $Gr(mn,k)\times Gr(mn,\ell)$ whenever $k+\ell<m+n-2$, or $k+\ell=m+n-2$ and $C^{k,\ell}_{m-1}\ne 0$.

\medskip
{\bf Conjecture:}\ %\begin{conjecture}
$A(m,n,k,\ell)$ is a full dimensional real semi-algebraic \emph{proper} subset when \eqref{eq-l3} holds.
%\end{conjecture}
\medskip

Here, the term \lq real semialgebraic\rq\ means that the set is determined by a
finite number of polynomial equations and polynomial inequalities
in real variables. It is obvious from the definition of $A(m,n,k,\ell)$ that this conjecture implies the converse of (iii) of Theorem \ref{cor-l1}.
We do not know how to prove this conjecture yet
except for the case when $m=2$ or $m=n=3$, for which we will
give explicit examples of pairs $(D,E)$ such that there is no nonzero
product vector $x\otimes y\in D$ with $\bar x\ot y\in E$.
We hope to get back to this conjecture in the future.

Now, we exhibit examples satisfying \eqref{eq-l3}.
For simplicity we use the notation
$$
(k,\ell)\lhd m\otimes n
$$
when the relation \eqref{eq-l3} holds. First of all, Proposition \ref{relation} tells us:
\begin{eqnarray*}
(k,k) &\lhd & 2\otimes 2k,\qquad k=1,2,\dots\\
\left(\textstyle{\binom{k+1}{2}},\ \textstyle{\binom{k+2}{2}} \right) &\lhd & 3\otimes k(k+2),\qquad k=1,2,\dots\\
(k,\ell) &\lhd & n\otimes n,\qquad k+\ell=2n-2,\ k \ {\text{\rm and}}\ \ell\ {\text{\rm are odd}}.\\
(k,k) &\lhd & m\otimes n,\qquad m+n=2k+2, \ m\ {\text{\rm and}}\ n\ {\text{\rm are even}}.
\end{eqnarray*}
Some more sequences of examples may be found:
$$
(2k,6k+1)\lhd 4k\otimes (4k+3),\qquad k=1,2,\dots
%\\
%(4k-1,4k+2)&\in &(2k+1)\otimes (6k+2),\qquad k=1,2,\dots
$$
for example.
In low dimensional cases with $m\times n<10$, we list up all cases satisfying \eqref{eq-l3} as follows:
$$
(1,1)\lhd 2\otimes 2,\qquad (2,2)\lhd 2\otimes 4,\qquad (1,3)\lhd 3\otimes 3.
$$

To get examples, we use the matrix notation rather than the tensor notation. We will use the notation $\{e_{i,j}\}$
for the standard matrix units.
We begin with the simplest case
$$
(1,1)\lhd 2\otimes 2.
$$
Let $D$ and $E$ be the orthogonal complements of $2\times 2$ matrices $P$ and $Q$, respectively. If one of
$P$ or $Q$ is of rank one then it is easy to see that there is a rank one matrix $xy^*$ satisfying
\begin{equation}\label{rankone}
xy^*\in D,\qquad \bar xy^*\in E.
\end{equation}
Indeed, if $Q=zw^*$ is of rank one, then take $x,y$ so that $y\perp w$ and $x\perp Py$. Next, we consider the case when
$$
P=\left(\begin{matrix}1&0\\0&t\end{matrix}\right),\quad
Q=\left(\begin{matrix}a&b\\c&d\end{matrix}\right).
$$
In this case, there exists no rank one matrix satisfying (\ref{rankone}) if and only if $\{Py,\overline{Qy}\}$ spans $\mathbb C^2$
for every $y\in \mathbb C^2$ if and only if
$$
\left(\begin{matrix} y_1 & \bar a\bar y_1+\bar b\bar y_2\\ ty_2 &  \bar c\bar y_1+\bar d\bar y_2\end{matrix}\right)
$$
is nonsingular for any $y=(y_1,y_2)$. This happens typically if $a=d=0$ and $b\, c\, t<0$. An extreme case occurs when
$P$ is the identity and $Q=e_{1,2}-e_{2,1}$. In this case $xy^*\perp P$ means that $x$ is orthogonal to $y$, and $\bar xy^*\perp Q$
means that $x$ and $y$ are parallel to each other.

A little variation of the above argument gives required examples in the case
$$
(k,k)\lhd 2\otimes 2k,\qquad k=1,2,\dots
$$
Let $(P_i,Q_i)$ be a pair of $2\times 2$ matrices such that there is no rank one matrix
$xy^*\in P_i^\perp$ such that $\bar xy^*\in Q_i^\perp$, for $i=1,2,\dots,k$. Let $\tilde P_i$
(respectively $\tilde Q_i$) be a $2k\times 2$ matrix whose $i$-th $2\times 2$ block is $P_i$ (respective $Q_i$)
with zeros in other blocks. If we put
\begin{equation}\label{2k_2-exam}
D=\{P_1,\dots, P_k\}^\perp,\qquad E=\{Q_1,\dots, Q_k\}^\perp
\end{equation}
then it is clear that there is no rank one matrix satisfying (\ref{rankone}).

Another variation of the above argument also gives an example in the case of
$$
(1,3)\lhd 3\otimes 3.
$$
To do this, put
\begin{equation}\label{ex33}
D=I^\perp,\quad
E=\{e_{1,2}-e_{2,1}, e_{2,3}-e_{3,2}, e_{3,1}-e_{1,3}\}^\perp,
\end{equation}
where $I$ denotes the identity matrix.
It is now clear that there is no rank one matrix $xy^*$ in $D$ such that $\bar xy^*\in E$. Indeed,
$xy^*\in D$ means that $x\perp y$, and $\bar xy^*\in E$ means that $x$ and $y$ are parallel to each other.
All of these examples show the following:

\begin{prop}
Suppose that $m=2$ or $m=n=3$, with $k+\ell=m+n-2$. Then
the condition {\rm (C)} holds if and
only if the condition {\rm (\ref{int-eq})} holds.
\end{prop}

Now, we examine whether there exists an edge state $A$ such that
\begin{equation}\label{edge-st}
{\mathcal R}A=D,\qquad {\mathcal R}(A^\tau)=E
\end{equation}
when $(D,E)$ is given by (\ref{ex33}).
To do this, we use the duality between the convex cone
$\mathbb T$ and the cone $\mathbb D$ consisting of all
decomposable positive linear maps, as was developed
in Ref. \onlinecite{eom-kye} and
\onlinecite{kye_decom}. If there exists $A\in\mathbb T$
satisfying (\ref{edge-st}) then the dual face
$A^\prime$ of the cone $\mathbb D$ must be
the convex hull of the set
\begin{equation}\label{xxxx}
\{\phi_V,\phi^W: V\in D^\perp, W\in E^\perp\},
\end{equation}
where $\phi_V(X)=VXV^*$ and $\phi^W(X)=WX^\ttt W^*$.
Now, we calculate the map
$$
\Phi=\phi_I+\phi^{e_{1,2}-e_{2,1}}+\phi^{e_{2,3}-e_{3,2}}+\phi^{e_{3,1}-e_{1,3}}
$$
directly, to get
$$
\Phi(e_{i,j})=
\begin{cases}
I,&\qquad i=j,\\
0,&\qquad i\neq j
\end{cases}
$$
for $i,j=1,2,3$. Therefore, we see that $\Phi$ is nothing but the trace map
$X\mapsto {\text{\rm tr}}(X)I$
which is a typical
interior point of the cone $\mathbb D$. This means that the convex hull of the set (\ref{xxxx}) is not a face, and so
we conclude that there is not an edge state with the property (\ref{edge-st}).

In the $2\otimes 2$ case, our examples never give rise to examples of edge states since $\mathbb V_1=\mathbb T$
by the work of Woronowicz\cite{woronowicz}.
This is also true for the example (\ref{2k_2-exam}) in the $2\otimes 2k$ case, as a variation of the $2\otimes 2$ case.

In the remainder of this note,
we consider the possible classification of low dimensional edge states by their types.
Note that Theorem \ref{cor-l1} gives us upper bounds of dimensions. Lower bounds are given in Ref. \onlinecite{hlvc} in which it was shown that
%\begin{equation}\label{sep}
$$
A\in\mathbb T,\  \dim{\mathcal R}(A)\le m\join n \ \Longrightarrow\ A\in\mathbb V_1,
%\end{equation}
$$
where $m\join n$ denotes the maximum of $m$ and $n$.
In the case of $2\otimes 4$, the possible types of edge states are
$$
(5,5),\quad (5,6),\quad(6,5),\quad (6,6).
$$
Note that the cases of $(5,7)$ and $(7,5)$ can be ruled out by Proposition \ref{relation} (i).
The first example of PPTES given by Woronowicz \cite{woronowicz} is turned to be an edge state of type $(5,5)$ in the
$2\otimes 4$ system.
This example has been modified in Ref. \onlinecite{p-horo} to get a one parameter family of the same type.
It was shown in Ref. \onlinecite{agkl}
that any $(5,5)$ edge state generates an extreme ray
of the cone $\mathbb T$, where
examples of edge states of type $(5,6)$ also were found.
It seems to be unknown whether there exists a $(6,6)$ edge state or not,
even though it was shown \cite{agkl} that there is no $(6,6)$
PPTES which generates an extreme ray.

In the $3\otimes 3$ case, possible types of edge states are
$$
(4,4),\quad (5,5),\quad (5,6),\quad (5,7),\quad (6,6),\quad (5,8),\quad (6,7),\quad (6,8),
$$
here we list up the cases $s\le t$ by the symmetry.
Note that we can rule out the case of $(7,7)$ by Proposition \ref{relation} (ii) or (iii).
The first example of PPTES in the $3\otimes 3$ case given by Choi \cite{choi82} is turned out to be an edge state
of type $(4,4)$. Other examples of edge states of this type were constructed
using orthogonal unextendible product bases \cite{bdmsst} and indecomposable
positive linear maps \cite{ha+kye}. In both cases, the images of the states are completely entangled.
In the latter case, the kernels of the edge states have six product vectors, which
are generic among $5$-dimensional subspaces of $M_{3\times 3}$. It was also shown \cite{ha+kye} that
the latter one generates an extreme ray
of the cone $\mathbb T$.
We refer to recent papers \onlinecite{chen0}, \onlinecite{chen}, \onlinecite {hhms}, \onlinecite{sko} and \onlinecite{slm} for detailed studies
for edge states of type $(4,4)$.

An example of a different type was firstly given by St\o rmer \cite{stormer82}, which is an edge state of type $(6,7)$.
One parameter family of PPTES in Ref. \onlinecite{p-horo} give us edge states of the same type.
Only known examples of edge states had been of types $(4,4)$ and $(6,7)$ until those of types
$(5,6)$, $(5,7)$ and $(5,8)$ were constructed in Ref. \onlinecite{ha-kye-2} using generalized Choi maps \cite{cho-kye-lee},
which are indecomposable positive.
Edge states of types $(5,5)$ and $(6,6)$ were found in Ref. \onlinecite{clarisse} and \onlinecite{ha-3} independently, which were also shown to
generate extreme rays in Ref. \onlinecite{kim-kye} and \onlinecite{ha-4}.
It seems to be still unknown whether there exists a $(6,8)$ edge state or not in the $3\otimes 3$ case.

In order to find entanglement independent from
PPT condition, we first have to characterize subspaces
without product vectors. For example, one may ask
the maximum dimension of subspaces without product vectors.
This question involves complex polynomials for which
standard techniques from algebraic geometry are available.
See Ref. \onlinecite{bhat}, \onlinecite{part},
\onlinecite{walgate} and \onlinecite{wall}
for this line of reaseach.
See also Ref. \onlinecite{arveson} and
\onlinecite{ruskai}
for measure theoretic approach.

On the other hand, the problem in this paper
involves complex polynomials with conjugates,
which are essentially real polynomials as is
mentioned in our Conjecture. This makes the problem
more difficult. Finally, it would be of
independent interest to
have the complete solution of the equation
(\ref{eq-l3}).

\medskip
{\bf Acknowledgement:}\
YHK was partially supported by NRFK 2010-0007786. SHK was partially supported by
NRFK 2011-0001250.
The authors are grateful to the authors of Ref.
\onlinecite{lein} for the valuable discussions on the product vectors.
%They are also grateful to Mary Beth Ruskai
%for bringing their attention to the Ref.
%\onlinecite{arveson} and \onlinecite{ruskai},
%together with valuable comments
%on the first draft of this note.

\medskip

%\nocite{...}
%\bibliography{aipsamp}% Produces the bibliography via BibTeX.

\end{document}